\documentclass[prl,aps,twocolumn,english,showpacs,superscriptaddress]{revtex4-1}
\usepackage[T1]{fontenc}
\usepackage[latin9]{inputenc}
\usepackage{color}
\usepackage{amsmath}
\usepackage{graphicx}
\usepackage[normalem]{ulem}
\usepackage{hyperref}
\usepackage{cleveref}

\usepackage{babel}

\usepackage{ulem}
\begin{document}

\title{Effect of Cherenkov radiation on localized states interaction}

\author{A. G. Vladimirov}
\email{vladimir@wias-berlin.de}
\affiliation{Weierstrass Institute for Applied Analysis and Stochastics, Mohrenstrasse 39, D-10117 Berlin, Germany}
\affiliation{Lobachevsky State University of Nizhni Novgorod, pr. Gagarina 23, Nizhni Novgorod, 603950, Russia}
\author{S. V. Gurevich}
\affiliation{Institute for Theoretical Physics, University of M\"unster, Wilhelm-Klemm-Str.\,9, D-48149 M\"unster, Germany}
\affiliation{Center for Nonlinear Science (CeNoS), University of M\"unster, Corrensstr.\,2, 48149 M\"unster, Germany}
\author{M. Tlidi}
\affiliation{Facult\'e   des Sciences, Universit\'e Libre de Bruxelles (U.L.B.), CP 231, 
Campus Plaine, B-1050 Bruxelles, Belgium}



\begin{abstract}
We study theoretically the interaction of temporal localized states in  all fiber cavities and microresonator-based optical frequency comb generators. We show that Cherenkov radiation emitted in the presence of third order dispersion breaks the symmetry of their interaction and greatly enlarges the interaction range thus facilitating the experimental observation of the soliton bound states.  Analytical derivation of the reduced equations governing slow time evolution of the positions of two interacting localized states in the Lugiato-Lefever model with third order dispersion term is performed.  Numerical solutions of the model equation are in close agreement with analytical predictions.  
\end{abstract}

\pacs{ 42.60.Da, 42.60.Fc, 42.65.Sf, 42.65.Tg, 05.45.Yv}
\maketitle
Frequency comb generation in microresonators has revolutionized such research disciplines as metrology and spectroscopy \cite{Kippenberg,Ferdous}. This due to the development of laser-based precision spectroscopy, including the optical frequency comb technique \cite{Hansch}. Driven optical microcavities widely used for the generation of optical frequency combs can be modeled by Lugiato-Lefever equation \cite{Lugiato1987} that possesses solutions in the form of localized structures also called cavity solitons (CSs) \cite{Coen_OL_13,Herr_NP_14}. Localized structures of the Lugiato-Lefever model have been theoretically predicted in \cite{Scorggie_csf94} and experimentally observed in \cite{LeoNat_pho_10}. In particular, temporal CSs manifest themselves in the form of short optical pulses propagating in the cavity.  The experimental evidence of temporal CSs interaction performed in \cite{LeoNat_pho_10} indicated that due to a very fast decay of the their tails, stable CS bound states are hardly observable. It has been also theoretically shown that when periodic perturbations are present \cite{Akhmediev_03,Turae12}, radiation of weakly decaying dispersive waves, i.e., so-called Cherenkov radiation emitted by CSs leads to a strong increase of their interaction range \cite{Akhmediev95,Cherenkov_17}. Experimental investigation of this radiation induced by the high order dispersion was carried out in \cite{Jang14,Wang17}.  Numerical studies on CSs bound states in the presence of high order dispersions have been reported  in  \cite{Oliver06,  Tli_Lendert, Skryabin_10,Bahloul_13,Bahloul_14,Parra-Rivas14,Pedro_17}.

In this letter, we  provide an analytical description of how two CSs can interact under the action of the Cherenkov radiation induced by high order dispersion. For this purpose, we use the paradigmatic Lugiato-Lefever model with the third order dispersion term. We derive the equations governing the time evolution of the position of two well-separated CSs interacting weakly via their exponentially decaying tails. We demonstrate that the presence of the third order dispersion term breaking the parity symmetry of the model equation leads to a great increase of the CS interaction range and affects strongly the nature of the CSs interaction. We show that the interference between the dispersive waves emitted by two interacting CSs produces an oscillating pattern responsible for the stabilization of the CS bound states. In particular, we show that when two CSs interact, one of them remains almost unaffected by the interaction force. On the contrary, the second interacting CS is strongly altered by the dispersive wave emitted by the first one. 

The LL model  with high order dispersion terms has been introduced in \cite{Tl_07}. In what follows, we consider only the second and third orders of dispersion. In this case the intracavity field is governed by the following dimensionless equation:
\begin{equation}
\frac{\partial E}{\partial T}=E_{in}-(1+i\theta)E+i d_2 \frac{\partial^{2}E}{\partial t^{2}}+d_{3}\frac{\partial^{3}E}{\partial t^{3}}+iE|E|^{2}.\label{eq:1}
\end{equation}
Here $E=E(t, T)$ is the  complex electric field envelope, $T$ is the slow time variable describing the number of round trips in the cavity and $t$ is the normalized retarded time variable (fast time). The parameter $E_{in}$ denotes  the normalized injected field amplitude, and $\theta$ is the normalized frequency detuning. Further, $d_{2}$ and $d_{3}$ are the second and the third-order dispersion terms, respectively. Without loss of generality we rescale $d_2$ to unity. The homogeneous stationary solution (HSS) of Eq. (\ref{eq:1}) is obtained from $E_{in}^{2}=I_0[1+(\theta-I_0)^{2}]$ with $I_0=|E_0|^{2}$.  For $\theta<\sqrt{3}$ ($\theta>\sqrt{3}$) the HSS is monostable (bistable) as a function of the input intensity. When $d_3=0$, Eq.~(1) supports both periodic \cite{Lugiato1987} and CS \cite{Scorggie_csf94} stationary states even in the monostable regime. 

\begin{figure}
\includegraphics[scale=0.25]{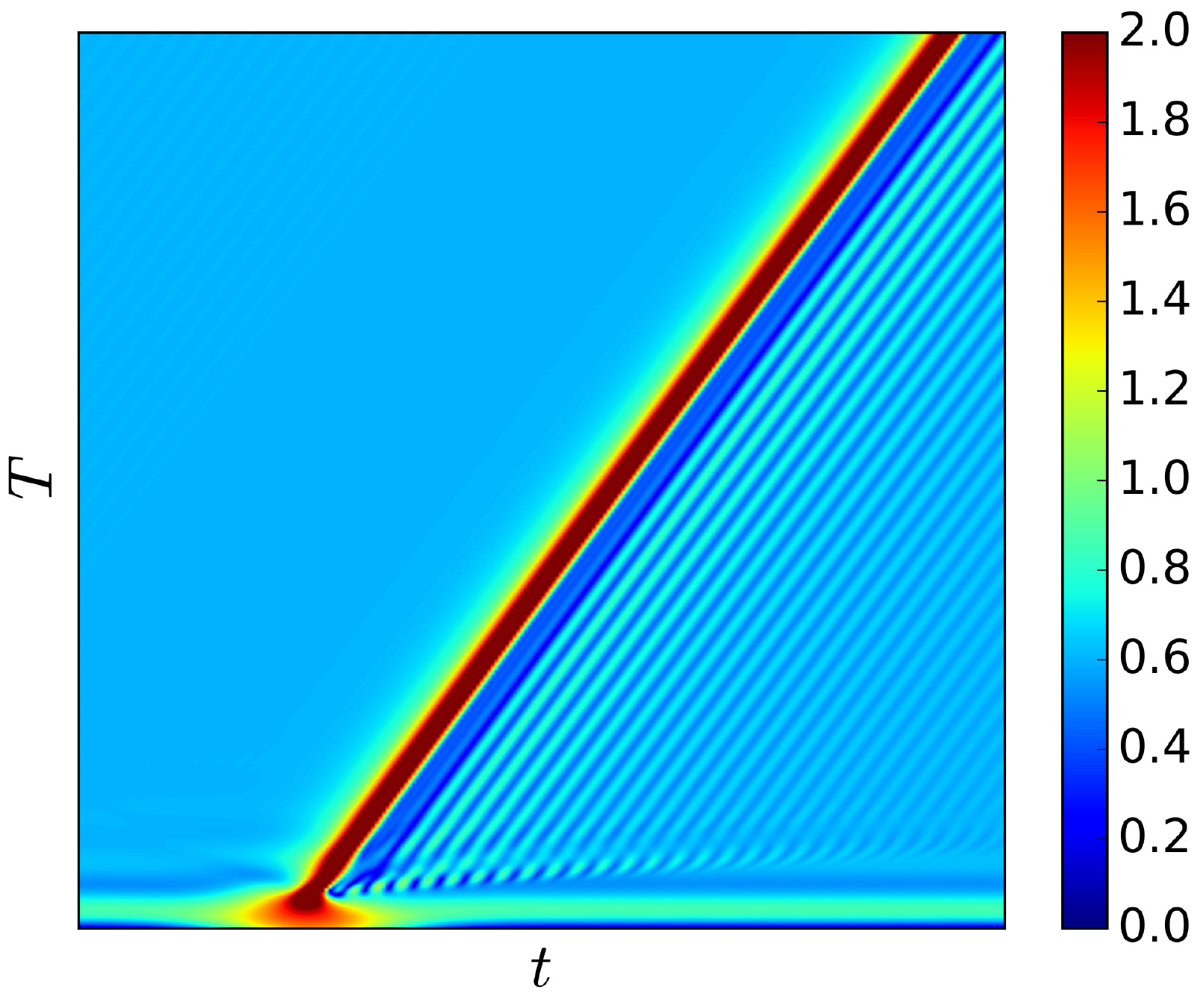}
\includegraphics[scale=0.25]{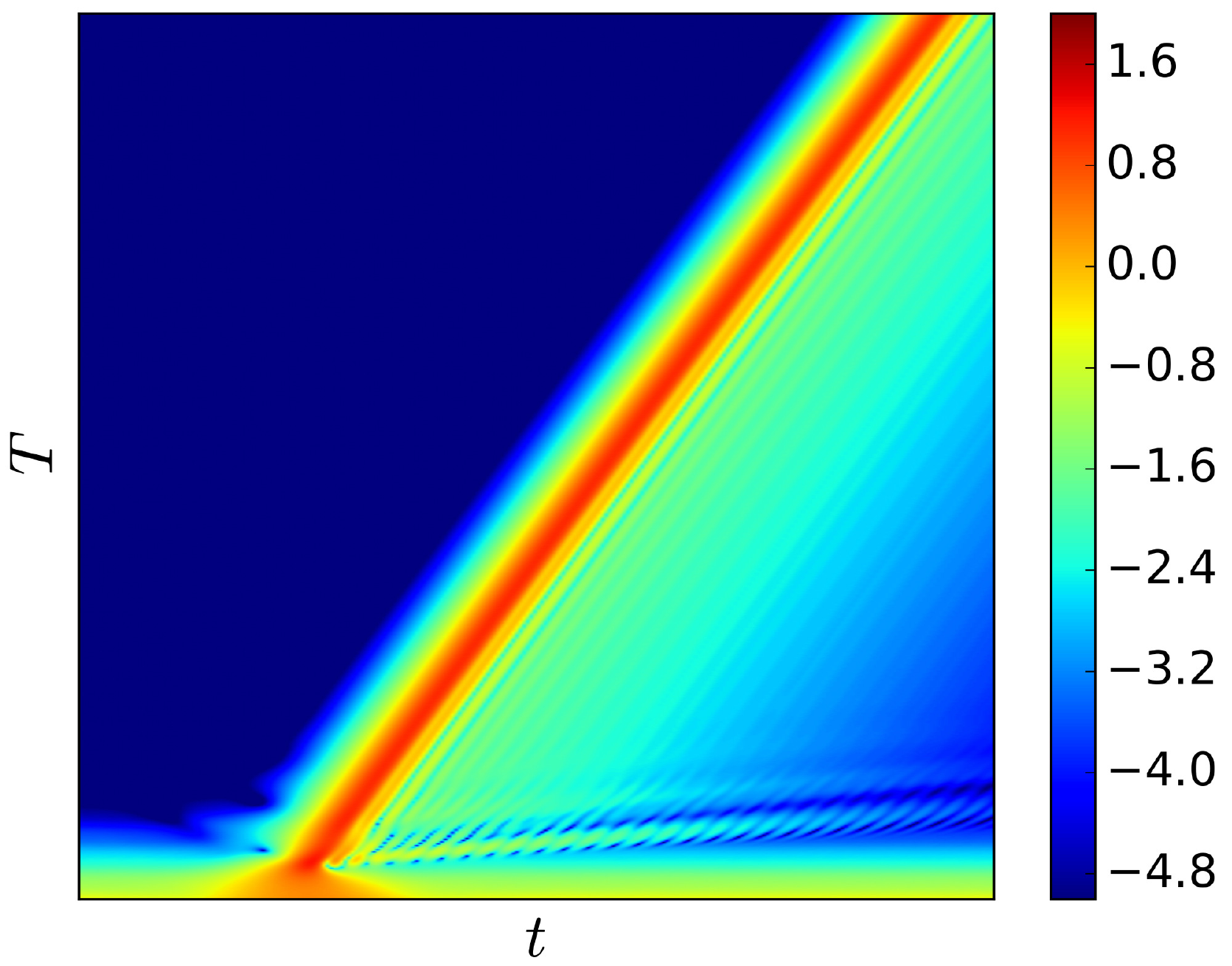}\\
\includegraphics[scale=0.3]{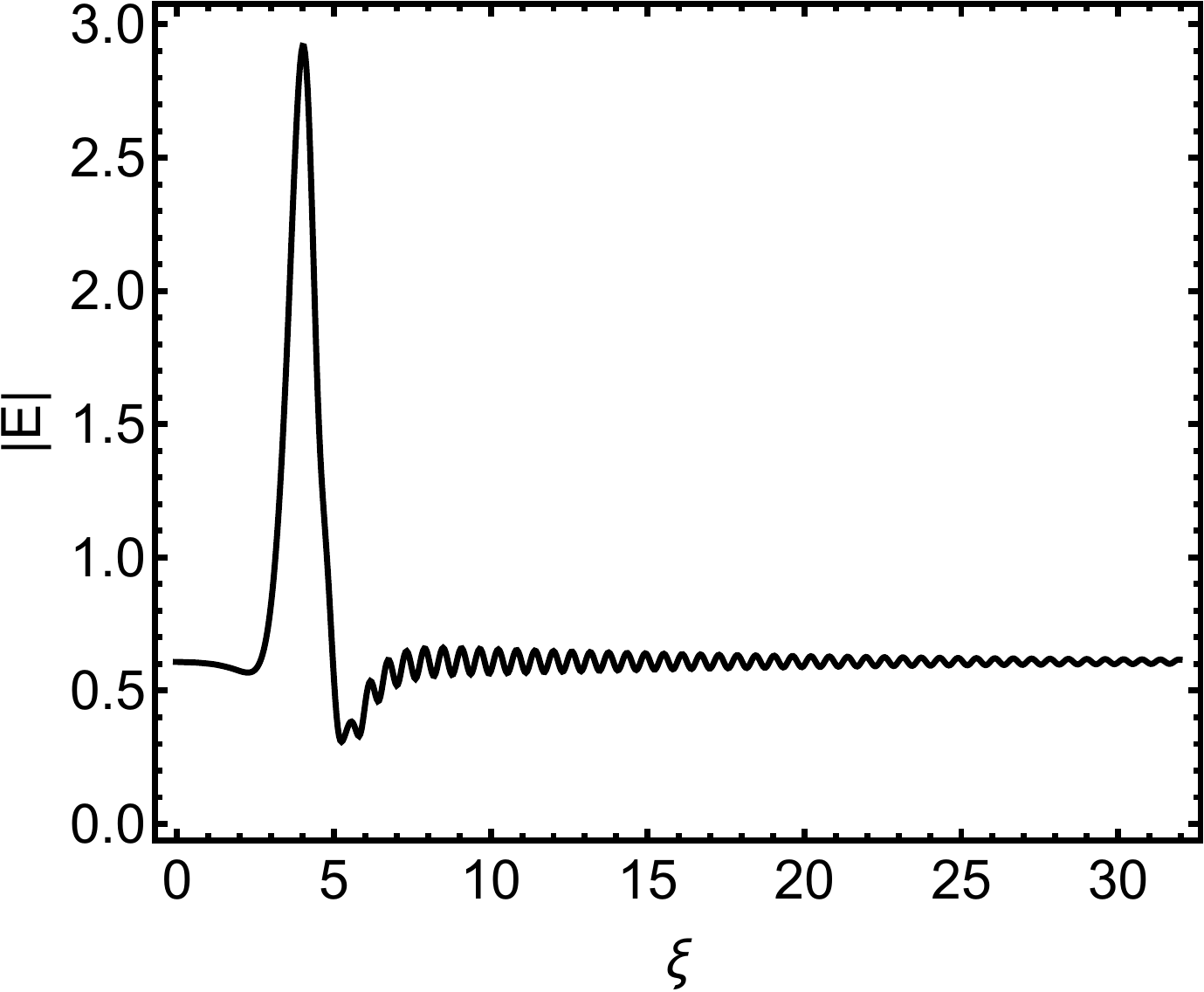} 
\includegraphics[scale=0.3]{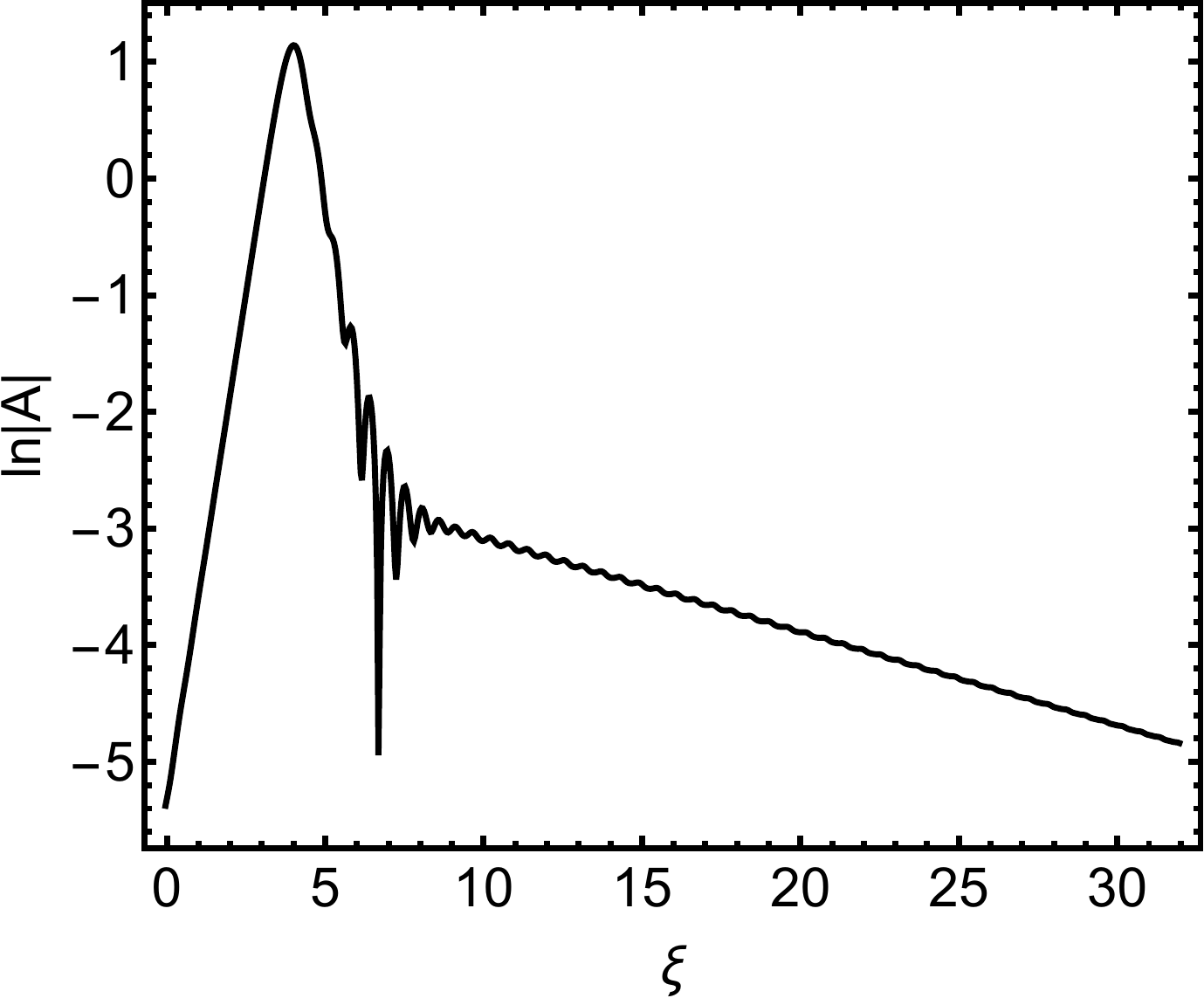}
\protect\caption{The amplitude $|E|$ of a CS calculated by numerical solution of Eq. (\ref{eq:1}) in linear scale (left) and the deviation $A(\xi)$ of the CS amplitude from the background in logarithmic scale (right). Top: CS formation in the $(t,T)$-plane ($d_{3}=0.2$). Bottom: CS moving uniformly with the velocity $v=0.50679$ ($d_3=0.1$). Other parameters are  $\theta=3.5$ and $E_{in}=2.0$. 
\label{fig:Soliton-amplitude}}
\end{figure}

When $d_3 \neq 0$, due to the breaking of the parity symmetry $t\to-t$, CS becomes asymmetric and starts to move uniformly with the velocity $v$ along the $t$-axis. An example of a moving CS obtained by direct numerical simulations of Eq. (\ref{eq:1}) with periodic boundary conditions is shown in Fig.~\ref{fig:Soliton-amplitude}, where the deviation of the CS amplitude from the HSS is defined as $A(\xi)=E(\xi)-E_0$ with  $\xi=t-vT$. It is seen from this figure that the inclusion of the third order dispersion induces an asymmetry in CS shape. The left (leading) CS tail decays very fast to the HSS $E=E_{0}$ as in the case when the third order dispersion is absent. By contrast, the right (trailing) tail contains a weakly decaying dispersive wave associated with the Cherenkov radiation \cite{Akhmediev95}. Note that the phase matching condition between the CS and the linear dispersive wave leads to a resonant wave amplification \cite{Akhmediev95,Skryabin_10} which is responsible for the appearance of this radiation.

\begin{figure}
\includegraphics[scale=0.6]{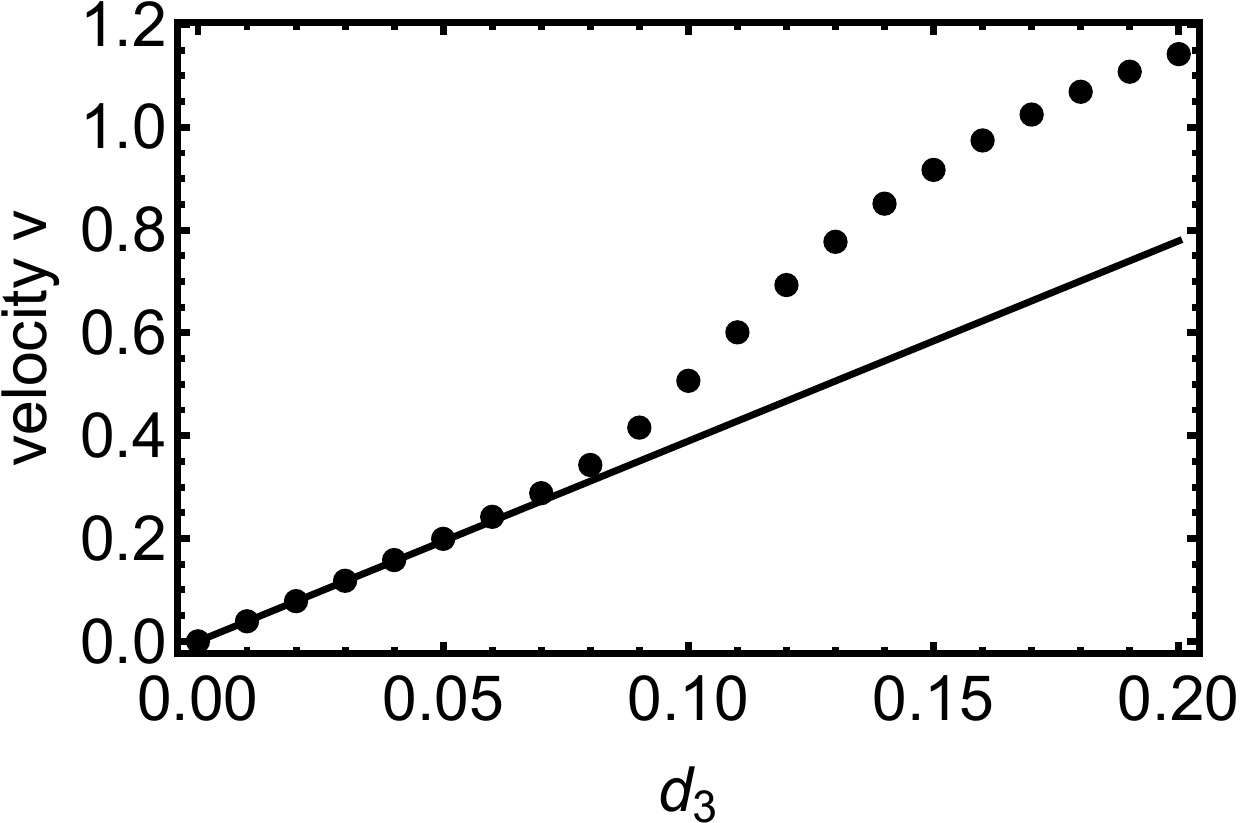}
\protect\caption{Soliton velocity $v$ vs. third order dispersion coefficient $d_{3}$. Solid line corresponds to the plot of the asymptotic formula (\ref{eq:v}) with numerically calculated $s=3.895$. Dots
indicate soliton velocities obtained by means of numerical solution of Eq.~(\ref{eq:1}). Parameter values are the same as in Fig.~\ref{fig:Soliton-amplitude} \label{fig:velocity}}
\end{figure}

The velocity $v$ of the CS can be estimated asymptotically at small $d_{3}$ using the multiple-scale techniques  
\begin{equation}
v\approx=-d_{3}s,\quad s=\intop_{-\infty}^{\infty}{\mathbf w}_{0}\cdot\frac{\partial^{3}{\mathbf a}_{0}}{\partial x^{3}}d\xi\left(\intop_{-\infty}^{\infty}{\mathbf w}_{0}\cdot {\mathbf a}_{0}d\xi\right)^{-1},\label{eq:v}
\end{equation}
where the index ``$0$'' indicates that both the CS solution ${\mathbf a}_{0}=\left(\mathrm{Re} A\,, \mathrm{Im} A\right)^T_{d_{3}=0}$ and the adjoint neutral mode ${\mathbf w}_{0}={\mathbf w}_{d_{3}=0}$  are evaluated
at $d_{3}=0$. The soliton velocity estimated using Eq. (\ref{eq:v}) and calculated by numerical solution of the model equation
(\ref{eq:1}) is shown in Fig.~\ref{fig:velocity}. It is seen that the asymptotic expression (\ref{eq:v}) with the numerically calculated coefficient $s=3.895$ agrees very well with the results of direct numerical simulation of Eq.~(\ref{eq:1}) for $d_{3}\le0.1$, where the CS velocity depends linearly on the third order dispersion coefficient. Notice that in the conservative limit where losses and injection are absent, one can obtain $s=\theta=3.5$ \cite{Akhmediev95}.
The CS shown in Fig. \ref{fig:Soliton-amplitude} is generated in regime where the system exhibits a bistable behavior. Let $E=E_0$ be the stable HSS with smallest field intensity $I_{0}=|E_0|^{2}$.  At large distances from the core the CS tails decay exponentially to this HSS. In order to characterize the asymptotic behavior of the CS tails, we substitute $E_{0}+\epsilon be^{\lambda\xi}$ into Eq. (\ref{eq:1}) and collect first order terms in the small parameter $\epsilon$. This yields the following characteristic equation: 
$
d_{3}^{2}\lambda^{6}+(1+d_{3}v)\lambda^{4}-2d_{3}\lambda^{3}+\lambda^{2}(4I_{0}+v^{2}-2\theta)-2v\lambda +\theta^{2}+1+3I_{0}^{2}-4\theta I_{0}=0$ for the eigenvalue $\lambda$. In the absence of third order dispersion, when $d_{3}=0$ and $v=0$, four  solutions of the characteristic equation are given by the expression $\lambda=\pm\sqrt{\theta-2I_{0}\pm\sqrt{I_{0}{}^{2}-1}}$. In the case when $I_{0}<1$ this expression gives  two pairs of complex conjugated eigenvalues $\pm\lambda_{0}$ and  $\pm\lambda_{0}^{*}$.  For small nonzero $d_3$ the eigenvalues $\pm\lambda_{0}$ and $\pm\lambda_{0}^{*}$ are transformed  into a pair of stable complex conjugated $\lambda_{1,2}$ and a pair of unstable complex conjugated (or real) eigenvalues, $\lambda_{5}$ and $\lambda_{6}$, located in small neighborhoods of $\pm\lambda_{0}$ and $\pm\lambda_{0}^{*}$ in the complex plane. More importantly, a pair of new eigenvalues, $\lambda_{3}$ and $\lambda_{4}=\lambda_{3}^{*}$ appears. In the limit of small third order dispersion $|d_{3}|\ll1$ the eigenvalues $\lambda_{3,4}$ can be written as 
$
\lambda_{3,4}=-d_{3}\mp i\left[\frac{1}{d_{3}}+d_{3}\left(\theta-2I_{0}-\nu\right)\right]+{\cal O}(d_{3}^{2}),
$
where we have neglected the term $v^{2}={\cal O}(d_{3}^{2})$. These new eigenvalues having small real and large imaginary parts are associated with the weakly decaying linear dispersive wave (Cherenkov radiation) emitted by CSs. As we will see below, they are responsible for the increase of the CS interaction range and formation of a large number of
bound states with large CS separations. In the anomalous dispersion regime, the dispersion coefficient $d_{3}$ is positive and the eigenvalues $\lambda_{3,4}$ have negative real parts. In this case the Cherenkov radiation appears at the trailing tail of the CS. At sufficiently large distances from the CS core this tail can be represented in asymptotic form 
\begin{equation}
A(\xi)\approx b_{1}e^{\lambda_{1}\xi}+b_{2}e^{\lambda_{2}\xi}+b_{3}e^{\lambda_{3}\xi}+b_{4}e^{\lambda_{4}\xi},\quad\xi\to+\infty,\label{eq:as2}
\end{equation}
where the coefficients $b_{3,4}$ can be considered as amplitudes of the Cherenkov radiation. Furthermore, linearizing Eq. (\ref{eq:as2}) at $A=0$ we
obtain $b_{1,4}=p_{1,4}b_{2,3}^{*}$ with 
\begin{equation}
p_{1,4}=\frac{E_{0}^{2}}{\theta-2|E_{0}|^{2}-i\kappa+iv\lambda_{1,4}-\lambda_{1,4}^{2}+id_{3}\lambda_{1,4}^{3}}.\label{eq:b14}
\end{equation}
For the parameter values of Fig.~\ref{fig:Soliton-amplitude} and $d_3=0.1$ numerical estimation of $b_{2,3}$ and $p_{1,4}$ gives $b_2=3.286 + 1.581 i $, $b_3 = -0.0678 + 0.0286 i $, $p_1=0.0221-0.0856 i$, and $p_4 =-0.001297-0.000895 i$.

\begin{figure}
\includegraphics[scale=0.6]{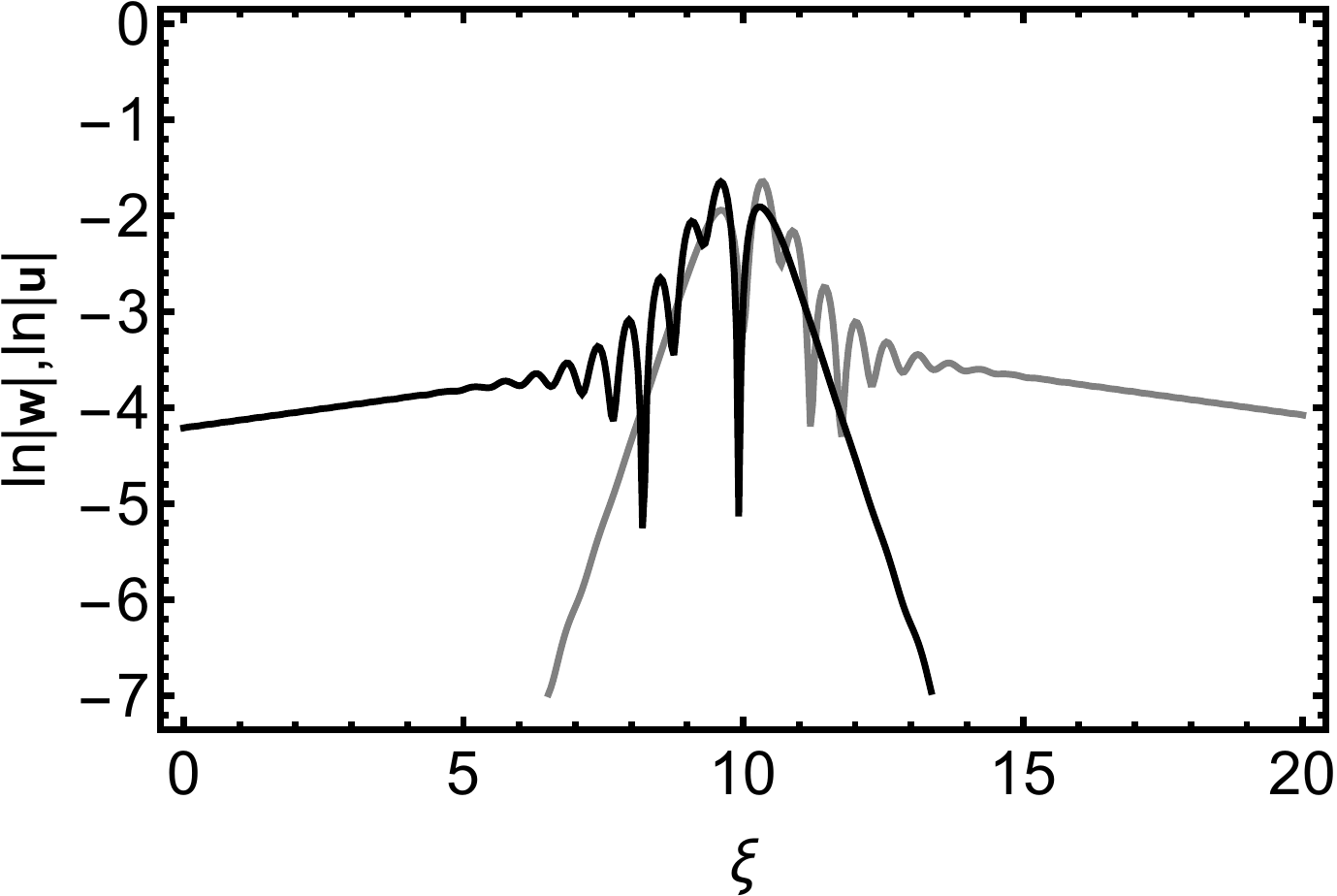}
\protect\caption{Neutral mode $|{\mathbf u}|$ (gray) and adjoint neutral mode $|{\mathbf w}|$ (black) in logarithmic scale calculated for $d_3=0.1$. Other parameters are the same as for Fig.~\ref{fig:Soliton-amplitude}. \label{fig:W}}
\end{figure}

It follows from Eq.~(\ref{eq:b14}) that $|p_{4}|={\cal O}(d_{3}^{2})$ in the limit $d_{3}\to0$, which means that small last term in (\ref{eq:as2}) can be omitted in the asymptotic analysis of the CS interaction. Therefore, since the eigenvalue $\lambda_3$ has small real part, at large positive $\xi$ the third term in Eq.~(\ref{eq:as2}) with the amplitude $b_3$ dominates in the weakly decaying and oscillating CS trailing tail. This coefficient is exponentially small in the limit $d_3\to 0$ and can be estimated analytically using the techniques similar to that described in the conservative limit \cite{Karpman93,Akhmediev95}. This is, however, beyond the scope of the present work. Stable eigenvalues $\lambda_{5,6}$ are responsible for the fast decay of the CS leading edge at negative $\xi\to-\infty$.
\begin{equation}
A(\xi)\approx b_{5}e^{\lambda_{5}\xi}+b_{6}e^{\lambda_{6}\xi},\quad\xi\to-\infty.\label{eq:as1}
\end{equation}
Numerical estimation gives the following values of the coefficients $b_{5,6}$: $b_5=0.111 -1.50 i$ and $b_6 =3.54+4.83 i$. Due to the translational invariance of Eq.~(\ref{eq:1}) along the $t$-direction, the linear operator $\hat{L}({\mathbf a})$ with ${\mathbf a}=\left(\mathrm{Re} A\,, \mathrm{Im} A\right)^T$ obtained by linearization of Eq.~(\ref{eq:1}) on the CS solution has zero eigenvalue corresponding to the so-called neutral translational eigenmode ${\mathbf u}=\partial_{\xi}\left(\mathrm{Re} A\ \mathrm{Im} A\right)^T$ satisfying the relation $\hat{L}({\mathbf a}){\mathbf u}=0$. In what follows, we will need also the neutral mode ${\mathbf w}=\left(\mathrm{Re} W\,, \mathrm{Im} W\right)^T$ of the linear operator $\hat{L}^{\dagger}({\mathbf a})$ adjoint to $\hat{L}({\mathbf a})$, which satisfies the relation $\hat{L}^{\dagger}({\mathbf a}){\mathbf w}=0$. The asymptotic behavior of the function $W$ defining the two components of the adjoint neutral mode $\mathbf w$ is given by the relations 
\begin{equation}
W(\xi)\approx c_{1}e^{-\lambda_{1}^{*}\xi}+c_{2}e^{-\lambda_{2}^{*}\xi}+c_{3}e^{-\lambda_{3}^{*}\xi}+c_{4}e^{-\lambda_{4}^{*}\xi},\quad\xi\to-\infty,\label{eq:as3}
\end{equation}
\begin{equation}
W(\xi)\approx c_{5}e^{-\lambda_{5}\xi}+c_{6}e^{-\lambda_{6}\xi},\quad\xi\to\infty,\label{eq:as4}
\end{equation}
with $c_{1,4}=-p_{1,4}^{*}c_{2,3}^{*}$ and the coefficients $p_{1,4}$ defined by Eq.~(\ref{eq:b14}). Numerical estimation of the coefficients $c_{2,3,5,6}$ yilds $c_2=-0.313+0.252 i$,  $c_3=-0.0152-0.0294 i$, $c_5=-0.185-0.0991 i$, and $c_6=0.245+0.456 i$. Similarly to $|b_4|\ll |b_3|$ the absolute value of the coefficient $c_4$ is much smaller than that of $c_3$. Hence, the term proportional to $c_4$ can be neglected in Eq.~(\ref{eq:as3}) when deriving the CS interaction equations. Absolute values of the neutral mode $|{\mathbf u}|=|\partial_{\xi}A|$ and  the adjoint neutral mode $|{\mathbf w}|$ are shown in Fig. \ref{fig:W} in logarithmic scale. From this figure we see that the neutral (adjoint neutral) mode has weakly decaying trailing (leading) tail.

\begin{figure}
\includegraphics[scale=0.5]{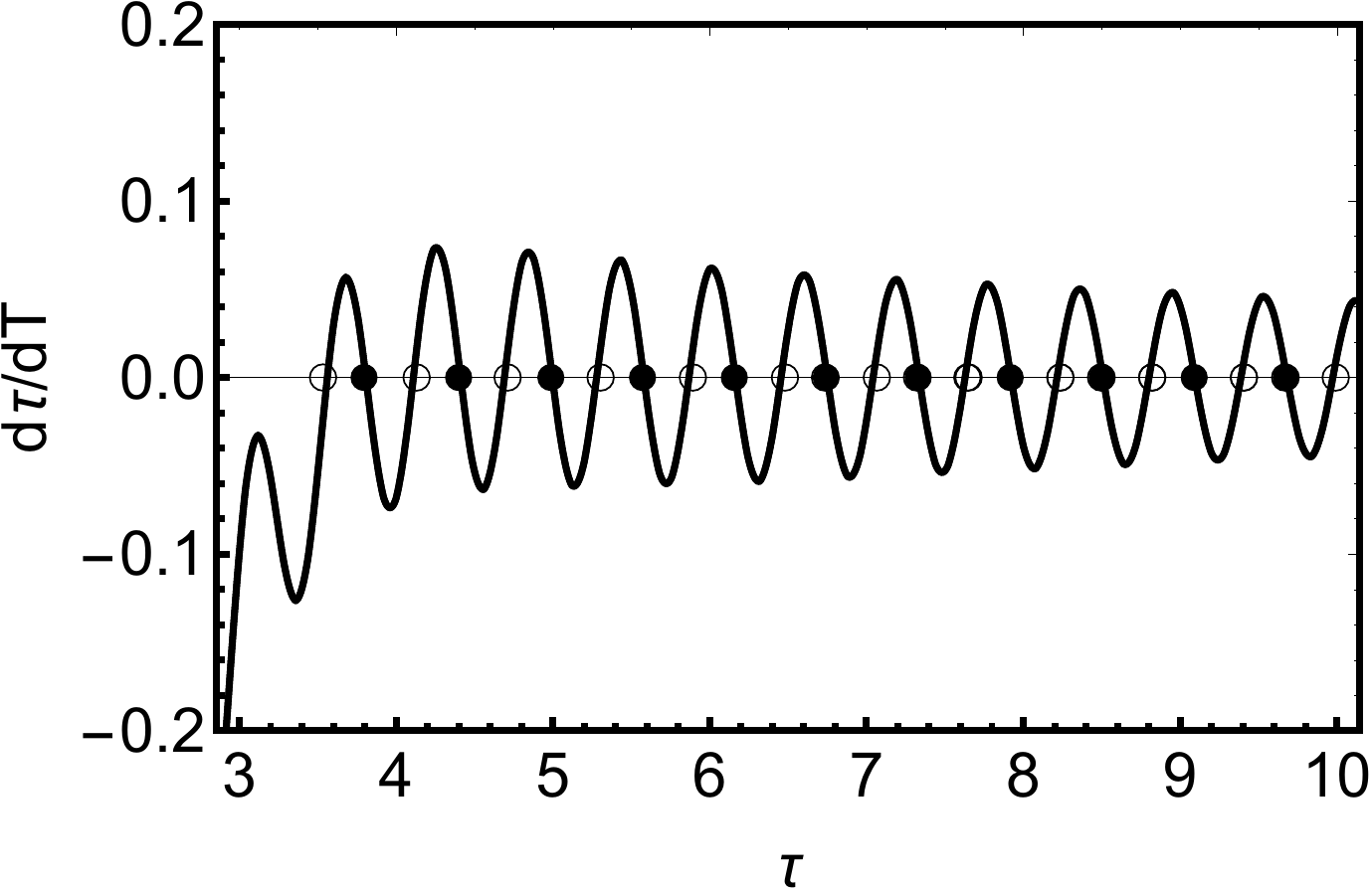} \\ \includegraphics[scale=0.5]{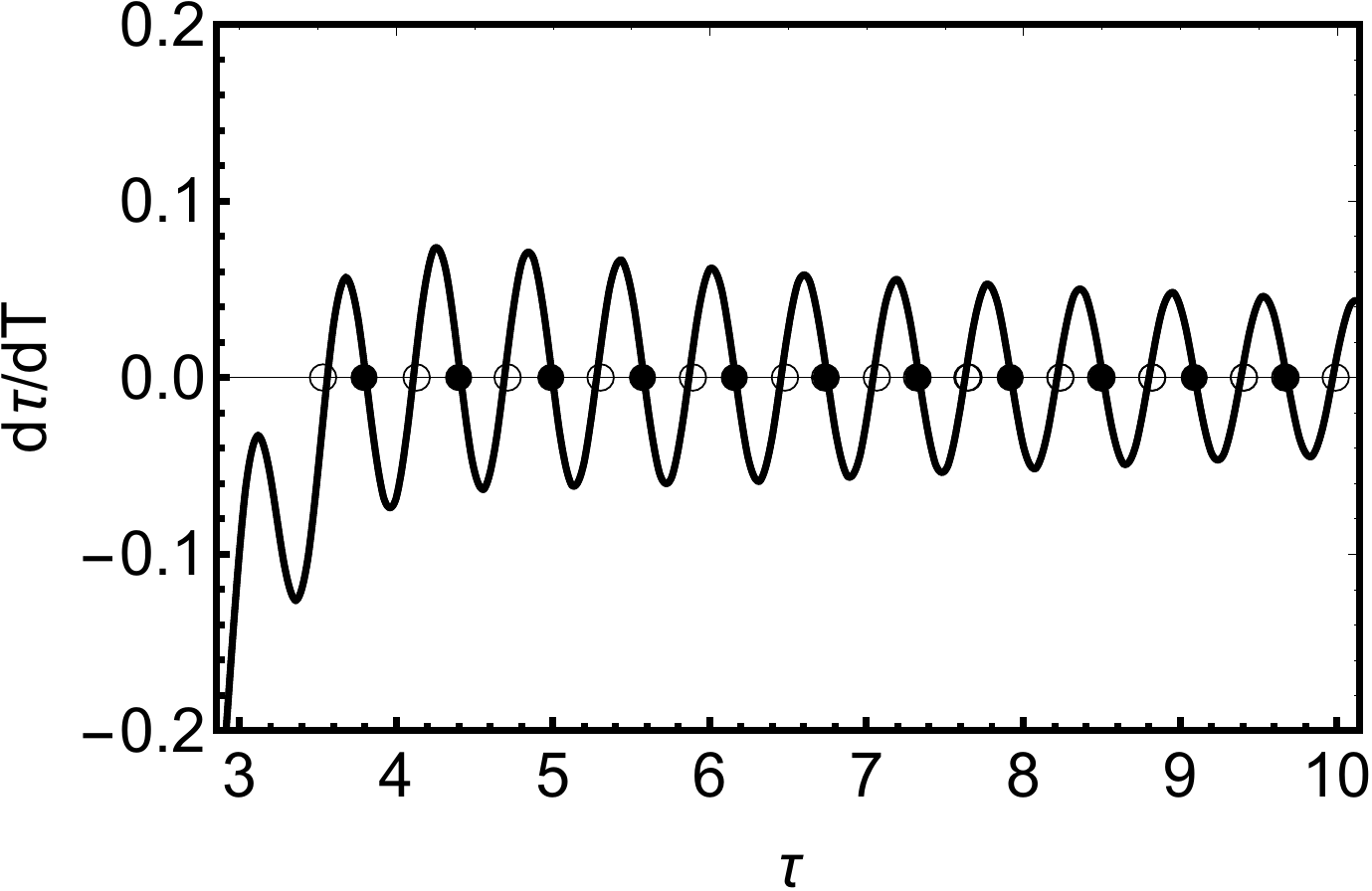}

\caption{Top: The dependence of CS velocities on their time separation. Unlike the velocity of the  first CS (black line), which is positive and fast decaying with the increase separation $\tau=\tau_2-\tau_1$, the velocity of the second CS (gray line) decays very slowly and oscillates fast as $\tau$ changes. Bottom: Difference of CS velocities as a function of their time separation. Zeros of this difference correspond to bound CS states. Numerically calculated CS time separations in the bound states are indicated by dots. Stable (unstable) bound states are shown by filled (empty) dots and correspond to decreasing (increasing) CS velocity difference. $d_3=0.1$, other parameters are the same as in Fig.~\ref{fig:Soliton-amplitude}.\label{fig:Interaction-forces}}
\end{figure}

\begin{figure}
\includegraphics[scale=0.30]{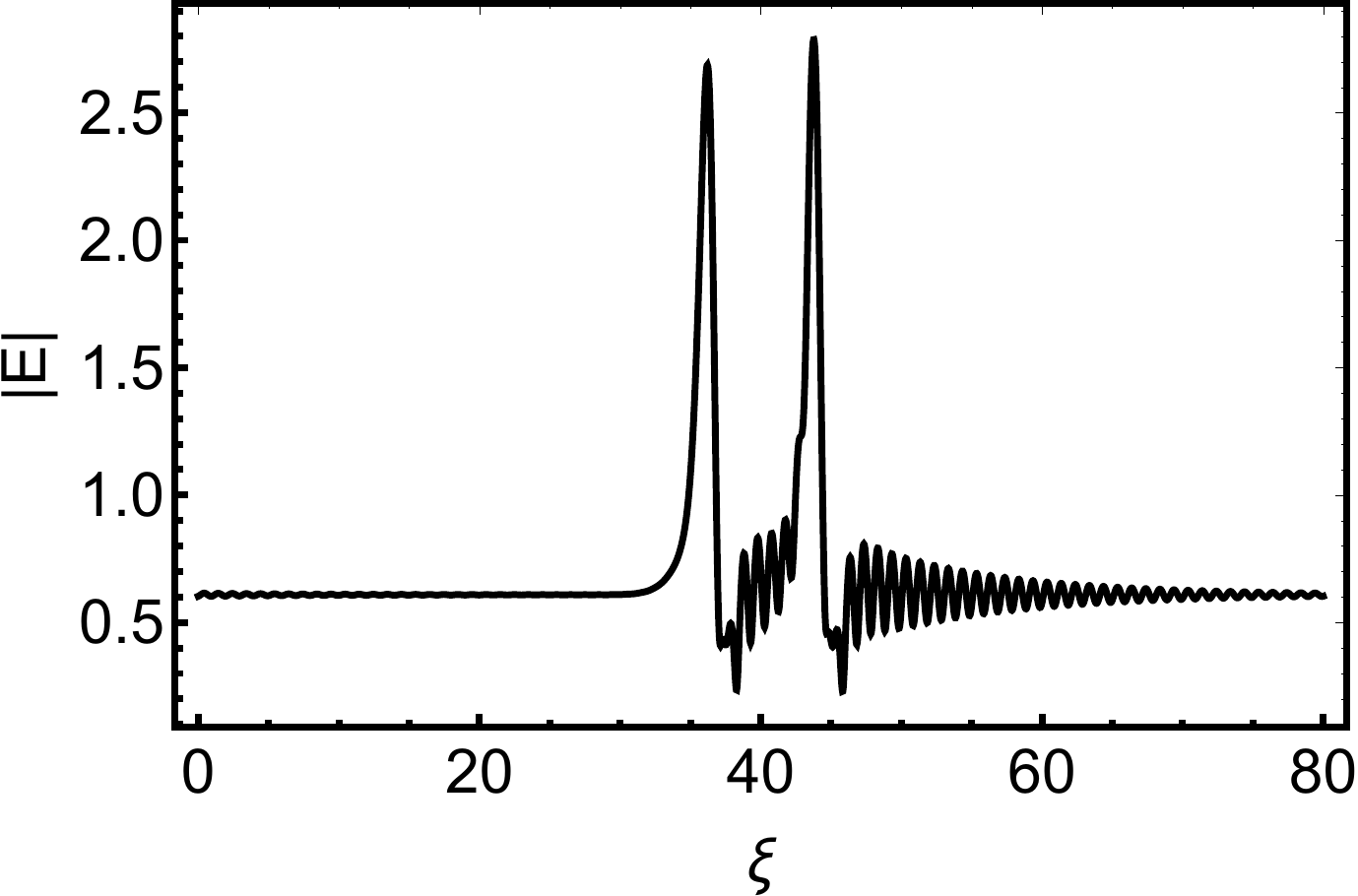}
\includegraphics[scale=0.29]{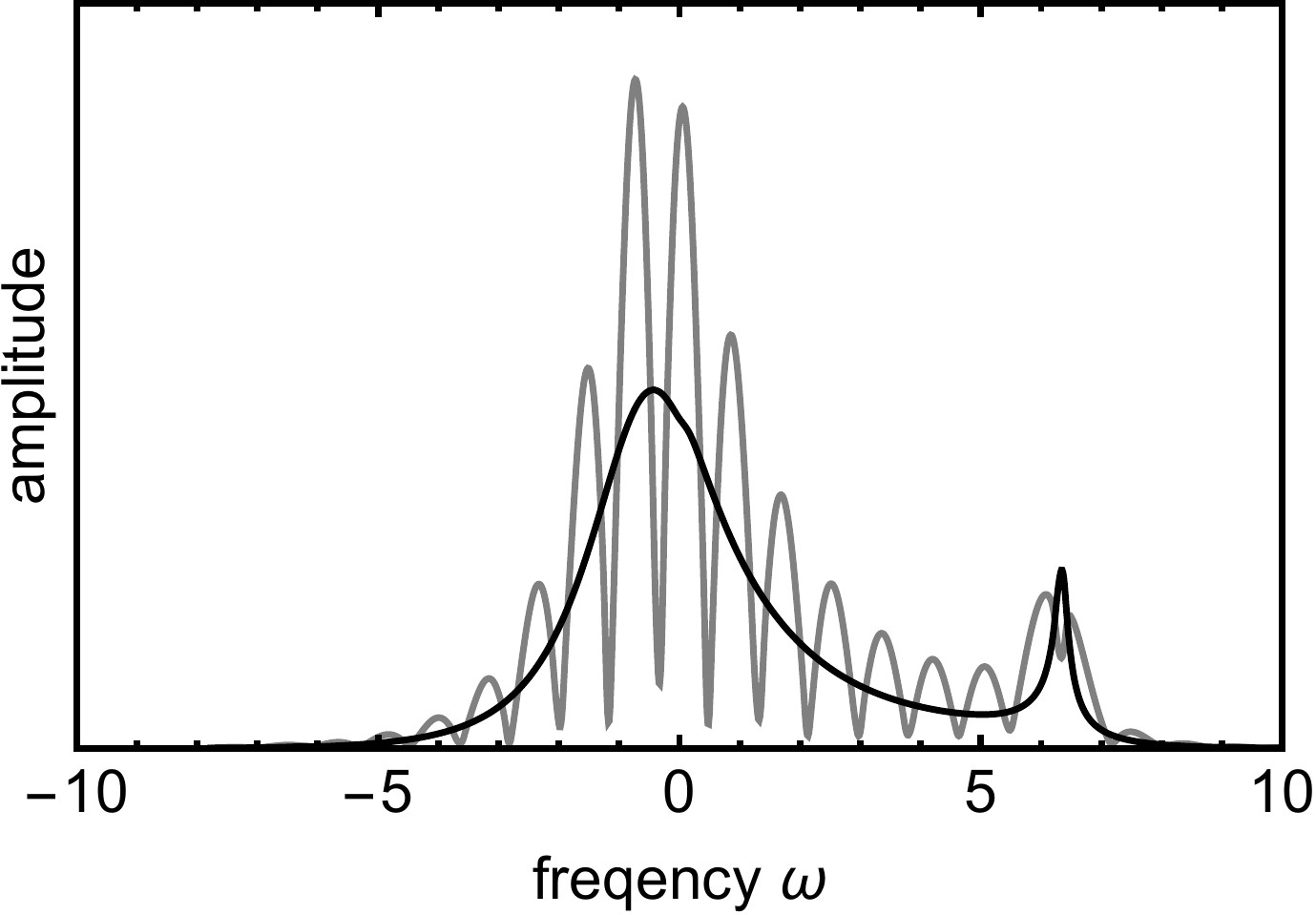}
\protect\caption{Left: Stable bound state of two solitons calculated for $d_3=0.2$. Left CS is almost unaffected by the interaction while the right one has larger peak power ans is much stronger modified by the interaction force. Note that for unstable bound states the peak power of the right CS is smaller that that of he left one.  Right: Frequency comb envelope for a solitary pulse (black) and pulse bound state shown in left panel (gray). The envelope modulation period of the bound state comb is determine by the time separation of the two pulses. Other parameters are the same as in Fig.~\ref{fig:Soliton-amplitude}.\label{fig:BS}}
\end{figure}

\begin{figure}
\includegraphics[scale=0.26]{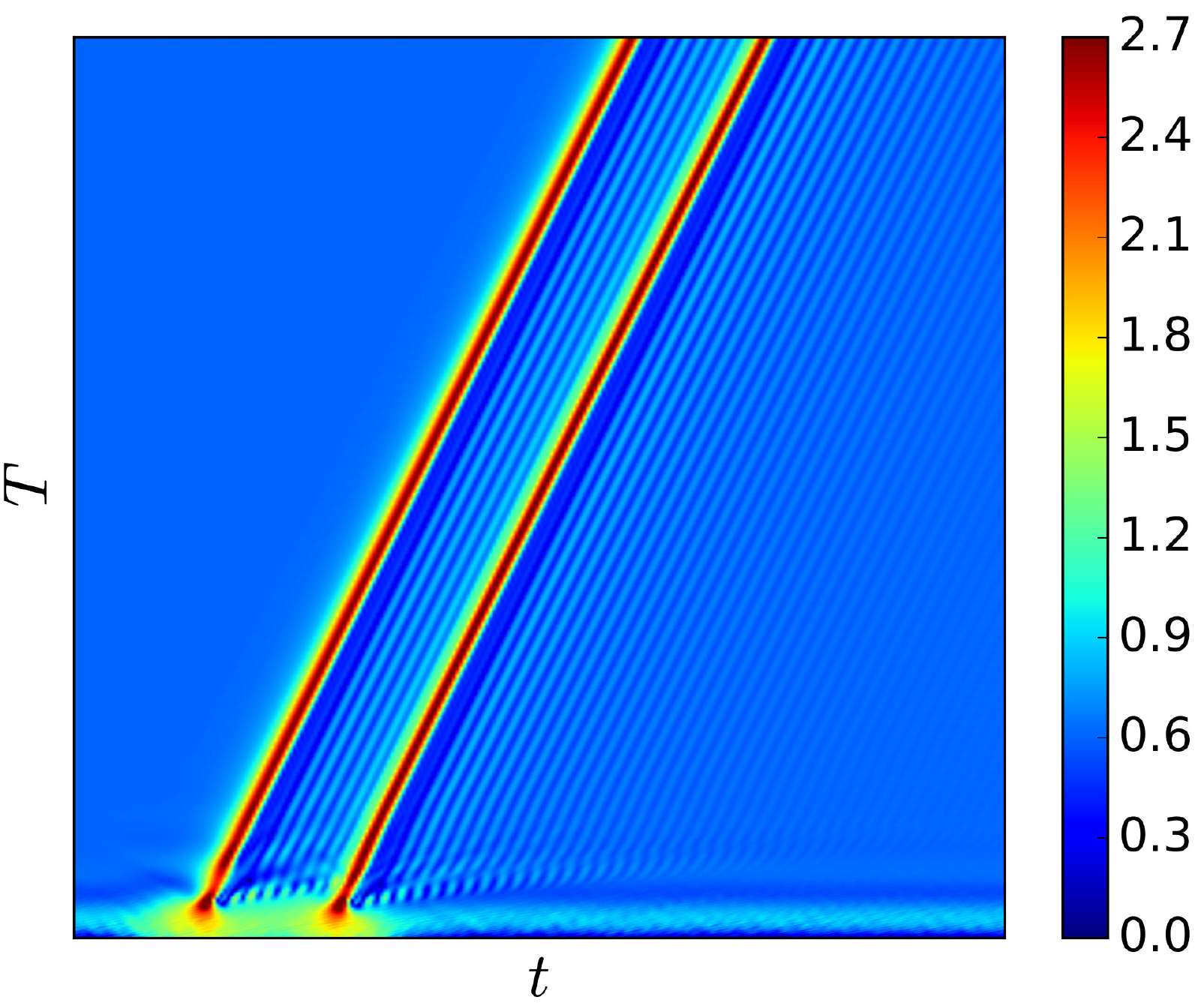}
\includegraphics[scale=0.26]{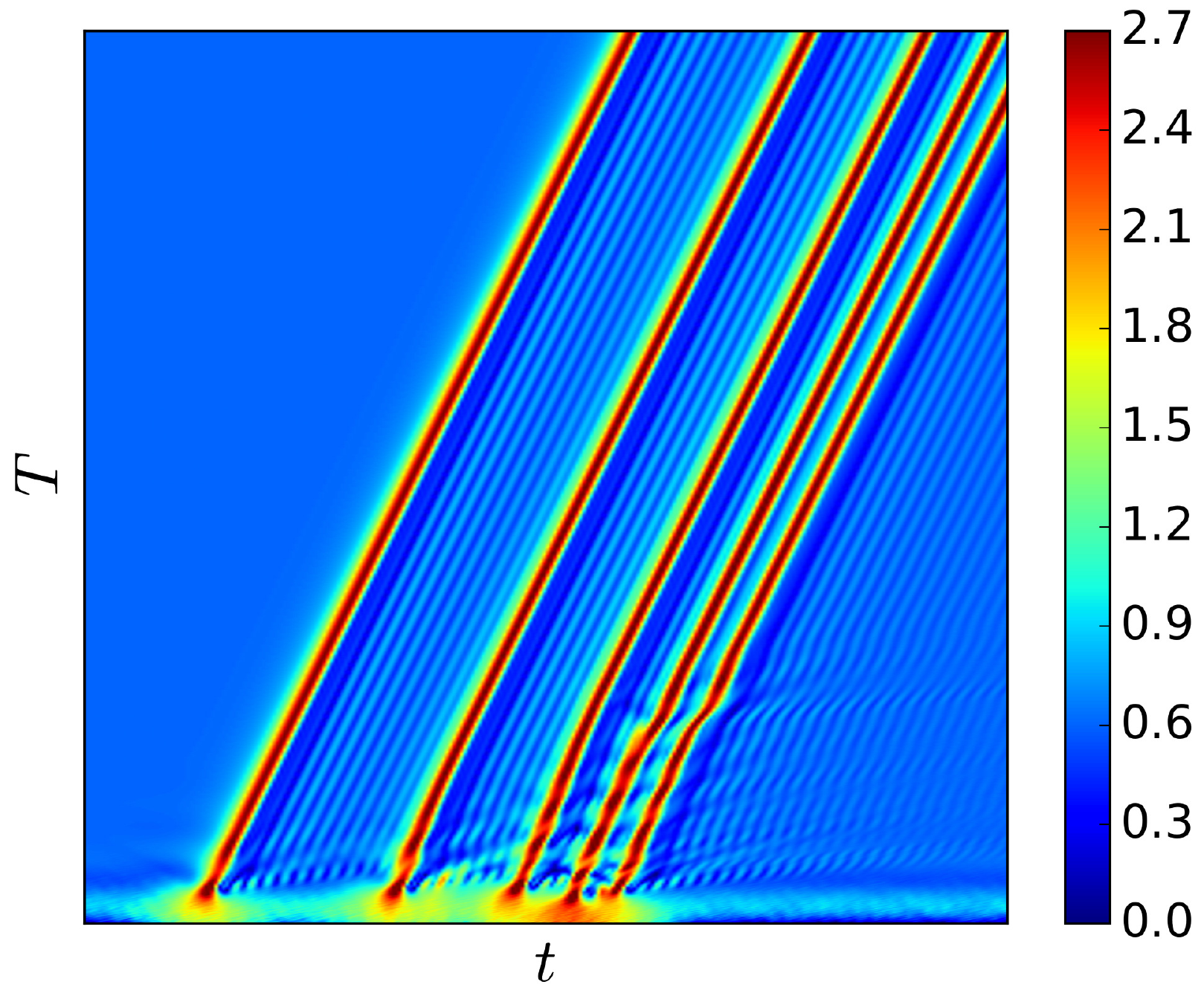}
\protect\caption{Formation of bound states of two (left) and five (right) CSs calculated for $d_3=0.2$. Other parameters are the same as in Fig.~\ref{fig:Soliton-amplitude}.\label{fig:two-five}}
\end{figure}

In order to derive the soliton interaction equations we use the Karpman-Solov'ev-Gorshkov-Ostrovsky approach and look for the solution of Eq.~(\ref{eq:1}) in the form of two weakly interacting CSs \cite{Vladimirov2002,TVP2003} 
\begin{equation}
E(\xi,t)=E_0+A_{1}+A_{2}+\delta A.\label{eq:Anzatz}
\end{equation}
Here, $A_{k}=A\left[\xi-\tau_{k}(T)\right]$ are unperturbed CS solutions with slowly changing coordinates along the $\xi$-axis, $d\tau_{1,2}/dT={\cal O}(\epsilon)$. The last term in the right hand side describes a small correction due
to the interaction, $\delta A={\cal O}(\epsilon)$, where the
parameter $\epsilon\ll1$ measures the weakness of the interaction.
Substituting (\ref{eq:Anzatz}) into the model equation (\ref{eq:1}) and collecting the terms of the first order in $\epsilon$ 
we get 
\begin{equation}
\hat{L}_{\Sigma}{\boldsymbol{\delta a}}=-\sum_{k=1}^2\frac{d\tau_{k}}{dT}{\mathbf u}_{k}+\hat{\mathbf f}_{\Sigma}.\label{eq:1storder}
\end{equation}
Here $\hat{L}_{\Sigma}=\hat{L}({\mathbf a}_{\Sigma})$, $\boldsymbol{\delta a}=\left(
\mathrm{Re}\delta A\,, \mathrm{Im}\delta A \right)^T$, ${\mathbf u}_{k}={\mathbf u}(\xi-\tau_{k})$ is the neutral mode of the $k$-th soliton,
and $\hat{f}_{\Sigma}=\left(\mathrm{Re}\hat{F}_{\Sigma}\,, \mathrm{Im}\hat{F}_{\Sigma}\right)^T$ with $\hat{F}_{\Sigma}=\hat{F}({\mathbf a}_{\Sigma})$ being the right hand side of (\ref{eq:1}) and ${\mathbf a}_{\Sigma}=\left[ \mathrm{Re} (A_{1}+A_{2})\,, \mathrm{Im} (A_1+A_2)\right]^T$.
The application of the solvability condition allows us to derive the velocities $d\tau_{1,2}/dt$ of the two interacting CS. Performing integration by parts in Eq.~(\ref{eq:1storder}), using asymptotic expressions (\ref{eq:as2})-(\ref{eq:as4}), and neglecting the terms proportional to small coefficients $b_{1}$, $b_{4}$, $c_{1}$, and $c_{4}$ we get:
\begin{equation}
\frac{d\tau_{2}}{dT}=\sum\limits_{n=2,3}\mathrm{Re}\left[b_{n}c_{n}^{*}\left(v+3d_{3}\lambda_{n}^{2}
+2i\lambda_{n}\right)e^{\lambda_{n}\tau}\right],\label{eq:int2}
\end{equation}
\begin{align}
 \frac{d\tau_{1}}{dT}=-\sum\limits_{n=5,6}\mathrm{Re}\left[b_{n}c_{n}^{*}
\left(v+3d_{3}\lambda_{n}^{2}
+2i\lambda_{n}\right)e^{-\lambda_{n}\tau}\right]+\nonumber\\
\mathrm{Re}\left[\left(b_{5}c_{6}^{*}+b_{6}c_{5}^{*}\right)
\left(v+\lambda_{56}^2+i\lambda_{56}-\lambda_5\lambda_6\right) e^{-\frac{\lambda_{56}\tau}{2}}\right],
\label{eq:int1}
\end{align}
where $\tau=\tau_{2}-\tau_{1}$ is the time separation of two CSs and $\lambda_{56}=\lambda_5+\lambda_6$. At small time separations the term with $n=2$ in the r.h.s. of (\ref{eq:int2}) and all the terms in the r.h.s. of Eq.~(\ref{eq:int1}) dominate in the interaction equations.  In particular, for $d_3=0.1$ when the eigenvalues $\lambda_{5,6}$ are real the two terms in (\ref{eq:int1}) are responsible for monotonous attraction of first CS to the second one. At larger CS separations, however, where the fast decaying r.h.s. of (\ref{eq:int1}) and the term with $n=2$ in (\ref{eq:int2}) become very small, the $n=3$ term in the r.h.s. of Eq.~(\ref{eq:int2}) related to the Cherenkov radiation becomes dominating. This slowly-decaying term oscillates fast with the CS time separation and it is responsible for bound state formation at large $\tau$. Thus at large CS separations Eqs.~(\ref{eq:int2}) and (\ref{eq:int1}) can be rewritten in the form clearly indicating the asymmetry of the soliton interaction:
\[
\frac{d\tau}{dT}=\frac{d\tau_{2}}{dT}=\mathrm{Re}\left[b_{3}c_{3}^{*}\left(3d_{3}\lambda_{3}^{2}+2i\lambda_{3}\right)e^{\lambda_{3}\tau}\right],\quad\frac{d\tau_{1}}{dT}\approx0.
\]
These equations predict the existence of an infinite countable set of equidistant stable CS bound states separated by unstable ones. They also indicate that at large $\tau$ the first CS is almost unaffected by the interaction, while the second CS moves in the potential created by the first one. The velocities $d\tau_{1,2}/dt$ of the two interacting solitons calculated using Eqs.~(\ref{eq:int2}) and (\ref{eq:int1}) with $d_3=0.1$ are shown in the top panel of Fig.~\ref{fig:Interaction-forces} as functions of the CS time separation $\tau$. The velocity of the first (left) soliton defined by the r.h.s. of Eq.~(\ref{eq:int1}) is a monotonous, always positive and fast decaying function of the CS time separation $\tau$. By contrast, the velocity of the second (right) soliton is negative only at relatively small $\tau$ and becomes slowly decaying and fast oscillating around zero at large $\tau$. This fast oscillating behavior is related to the Cherenkov radiation and described by the $n=3$ term in the r.h.s. of Eq.~(\ref{eq:int2}). It is responsible for the formation of CS bond states at sufficiently large time separations $\tau$. In order to find these states, we plot the difference of the CS velocities $d\tau/dt$  as a function of $\tau$ in the bottom panel of Fig.~\ref{fig:Interaction-forces}. Zeros of $d\tau/dt$ correspond to the fixed points of the CSs interaction equations. Stable (unstable) CSs bound states calculated by direct numerical solution of the model equation (\ref{eq:1}) are indicated  by filled (empty) dots in this figure. It is seen that they are in a good agreement with the results of the asymptotic analysis. Furthermore, a stable bound state of two CS and the corresponding frequency comb are shown in Fig.~\ref{fig:BS}. A ``space-time'' diagram in the ($T,\, t$) plane illustrating the formation of two-soliton and five-soliton bound states with different distances is shown in Fig.~\ref{fig:two-five}(a,\,b). 

\begin{figure}\end{figure}

To conclude, we have investigated the effect of Cherenkov radiation on the CS interaction in the generalized Lugiato-Lefever model with the third order dispersion term, which is widely used to describe frequency comb generation in optical microresonators and CS formation in fiber cavities. We have developed an analytical asymptotic theory of the CS interaction. The results of numerical simulation of the model equation are in good agreement with analytical predictions. We have shown that the third order dispersion greatly enlarges the CS interaction range and makes the interaction very asymmetric. This allows for the stabilization of large number of bounded states formed by CSs. As was mentioned above, in the absence of the third order dispersion, bound states are hardly observable experimentally due to rather fast decay and slow oscillation of the CS tail \cite{LeoNat_pho_10}. That is, considering the system operating close to the zero dispersion wavelength regime where the third order dispersion comes into play, one can facilitate experimental observation of the CS bound states.

\bigskip
 
 A. G. V. acknowledges the support of SFB 787, project B5 of the DFG and the Grant No. 14-41-00044 of the Russian Scientific Foundation. S. V. G. acknowledges the support of Center for Nonlinear Science (CeNoS) of the University of M\"{u}nster.  M. T. thanks the Interuniversity Attraction Poles program of the Belgian Science Policy Office under the grant IAP P7-35. M. T. received support from 
the Fonds National de la Recherche Scientifique (Belgium).

\end{document}